\documentclass[a4paper, 10pt, onecolumn]{article}      % Use this line for a4
\usepackage{graphics} % for pdf, bitmapped graphics files
\usepackage{epsfig} % for postscript graphics files
\usepackage{mathptmx} % assumes new font selection scheme installed
\usepackage{times} % assumes new font selection scheme installed
\usepackage{amsmath} % assumes amsmath package installed
\usepackage{amssymb}  % assumes amsmath package installed
\usepackage{theorem}
\newtheorem{theorem}{Theorem}
\newtheorem{remark}{Remark}

\title{\LARGE \bf
Multiple time-delays system modeling and control for router management
}

\author{Yassine Ariba$^{\ast\dag}$, Fr\'ed\'eric Gouaisbaut$^{\ast\dag}$ and Yann Labit$^{\ast}$% <-this % stops a space
\thanks{ Universit\'e de Toulouse; UPS, 118 Route de Narbonne, F-31062 Toulouse, France.}%
\thanks{ LAAS; CNRS;  7, avenue du Colonel Roche, F-31077 Toulouse, France.
        {\tt\small \{yariba,fgouaisb,ylabit\}@laas.fr}}}

\date{October 2008}

\begin{document}

\maketitle
\thispagestyle{empty}
\pagestyle{empty}

%%%%%%%%%%%%%%%%%%%%%%%%%%%%%%%%%%%%%%%%%%%%%%%%%%%%%%%%%%%%%%%%%%%%%%%%%%%%%%%%
\begin{abstract}
This paper investigates the overload problem of a single congested router in TCP ({\it Transmission Control Protocol}) networks. To cope with the congestion phenomenon, we design a feedback control based on a multiple time-delays model of the set TCP/AQM  ({\it Active Queue Management}). Indeed, using robust control tools, especially in the quadratic separation framework, the TCP/AQM model is rewritten as an interconnected system and a structured state feedback is constructed to stabilize the network variables. Finally, we illustrate the proposed methodology with a numerical example and simulations using NS-2 \cite{Fal02} simulator.

\end{abstract}

%%%%%%%%%%%%%%%%%%%%%%%%%%%%%%%%%%%%%%%%%%%%%%%%%%%%%%%%%%%%%%%%%%%%%%%%%%%%%%%%
\section{INTRODUCTION}
~\indent In IP networks, active queue management (AQM), embedded in router, reports to TCP sources its processing load. The objective is to manage the buffer utilization as well as the queueing delay.   This has motivated a huge amount of work aiming at understanding the congestion phenomenon and achieving better performances in terms of {\it Quality of Service} (QoS). As a matter of fact, there has been a growing recognition that the network itself must participate in congestion control and ressource management \cite{Bra98}, \cite{Le03}.~\\~\indent The AQM principle consists in dropping (or marking when ECN \cite{Ram99} option is enabled) some packets before buffer saturates. Hence, following the {\it Additive-Increase Multiplicative-Decrease} (AIMD) behavior of TCP, sources reduce their congestion window size avoiding then the full saturation of the router. Basically, AQM support TCP for congestion avoidance and feedback to the latter when traffic is too heavy. Indeed, an AQM drops/marks incoming packet with a given probability related to a congestion index (such as queue length or delays) allowing then a kind of control on the buffer occupation at routers. Various mechanisms have been proposed in the network community for the development of AQM such as Random Early Detection (RED) \cite{Flo93}, Random Early Marking (REM) \cite{Ath00},  Adaptive Virtual Queue (AVQ) \cite{Kun01} and many others \cite{Ryu04}. Their performances have been evaluated \cite{Fir00}, \cite{Ryu04} and empirical studies \cite{Le03} have shown the effectiveness of these algorithms. ~\\~\indent As it has been highlighted in the litterature (see for example \cite{Low02}, \cite{Hol02} and reference therein), AQM acts as a controller supporting TCP for congestion control and can be reformulated as a feedback control problem. Then, a significant research has been devoted to the use of control theory to develop more efficient AQM. Using dynamical model developed by \cite{Mis00}, some P ({\it Proportional}), PI ({\it Proportional Integral}) \cite{Hol02} have been designed. In the same framework, other tools have been used to extend this preliminary work such as a PID controller \cite{Ryu04}, \cite{Fan03} or robust control \cite{Que04}. However, most of these papers do not take into account the delay and ensure the stability in closed loop for all delays which could be very conservative in practice.\\~\indent The study of congestion problem in time delay systems framework is not new and has been successfully exploited. Nevertheless, most of these works have been dedicated to the stability analysis of networks composed of homogeneous sources (see for example \cite{Lab07b}, \cite{Pap04}, \cite{Chen07}, \cite{Man04} and \cite{Wan03}). In this paper, networks with heterogeneous sources are considered introducing then several delays and increasing the model complexity. Regarding the AQM design problem, some works have already been done in \cite{Kim06} and \cite{Han04}. In \cite{Han04}, the construction of the AQM required to invoke the Generalized Nyquist Theorem and \cite{Kim06} provides a delay dependent state feedback involving delay compensations with a memory feedback control. Or even in \cite{Awe07}, delays are replaced by a Pad\'e approximation which is known to be not so accurate. All these latter methodologies are interesting in theory but sorely suitable in practice. While these latter studies have considered the simplified model of TCP/AQM from \cite{Mis00}, we use in this contribution a more accurate model presented in \cite{Low02}. Indeed, contrary to \cite{Kim06}, \cite{Awe07} and \cite{Han04}, both forward and backward delays are taken into account (that is, we do not neglect forward delays). Then, congestion control of networks consisting in heterogeneous TCP sources is transformed into a stabilization problem for multiple time delays systems. Using a robust analysis framework and especially quadratic separation approach developed for time-delay systems by \cite{Gou06a}, a stabilizing AQM is designed. The proposed control mechanism enables QoS in terms of RTT ({\it Round Trip Time}) and delay jitter which are relevant features for streaming and real-time applications over IP networks. Note that the approach employed in this paper allows the formulation of the problem into matrix inequalities \cite{Boy94} that provide systematic stability condition, easy to test.\\~\indent The paper is organized as follows. The second part presents the mathematical model of a network composed of a single router and several heterogeneous sources supporting TCP. Section III is dedicated to the design of the AQM ensuring the stabilization of TCP. Section IV presents a numerical example and simulation results using NS-2.

%~\indent {\it Notations:} For
%two symmetric matrices, $A$ and $B$, $A>$ ($\geq$) $B$ means that
%$A-B$ is (semi-) positive definite. $A^T$ denotes the transpose of
%$A$. $\sf{1}_n$ and $\sf{0}_{m\times n}$ denote respectively the
%identity matrix of size $n$ and null matrix of size $m\times n$. If
%the context allows it, the dimensions of these matrices are often
%omitted.

\section{NETWORK DYNAMICS}

\begin{figure}
%    \begin{minipage}[c]{.50\linewidth}
       \centerline{\includegraphics[width=6cm,height=.19 \textheight]{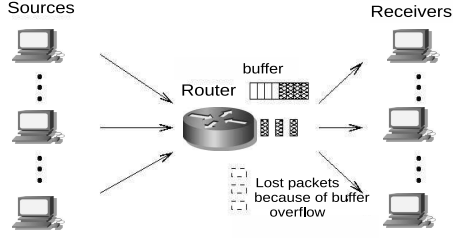}}
       \caption{Network topology}
       \label{topologie}
       \end{figure}

In this paper, we consider a network consisting of a single router and $N$ heterogeneous TCP sources. By heterogeneous, we mean that each source is linked to the router with different propagation times (see Figure \ref{topologie}). Since the bottleneck is shared by $N$ flows, TCP applies the congestion avoidance algorithm to cope with the network saturation \cite{Jac88}.
%    \end{minipage}
%    \hfill
%    \begin{minipage}[c]{.40\linewidth}
\begin{figure}
        \centerline{\includegraphics[height=.17 \textheight]{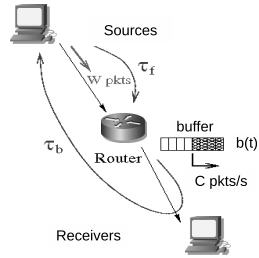}}
       \caption{A single connection}
       \label{topologie4}
%    \end{minipage}
\end{figure}

\begin{figure}
       \centerline{\includegraphics[width=6cm,height=.15 \textheight]{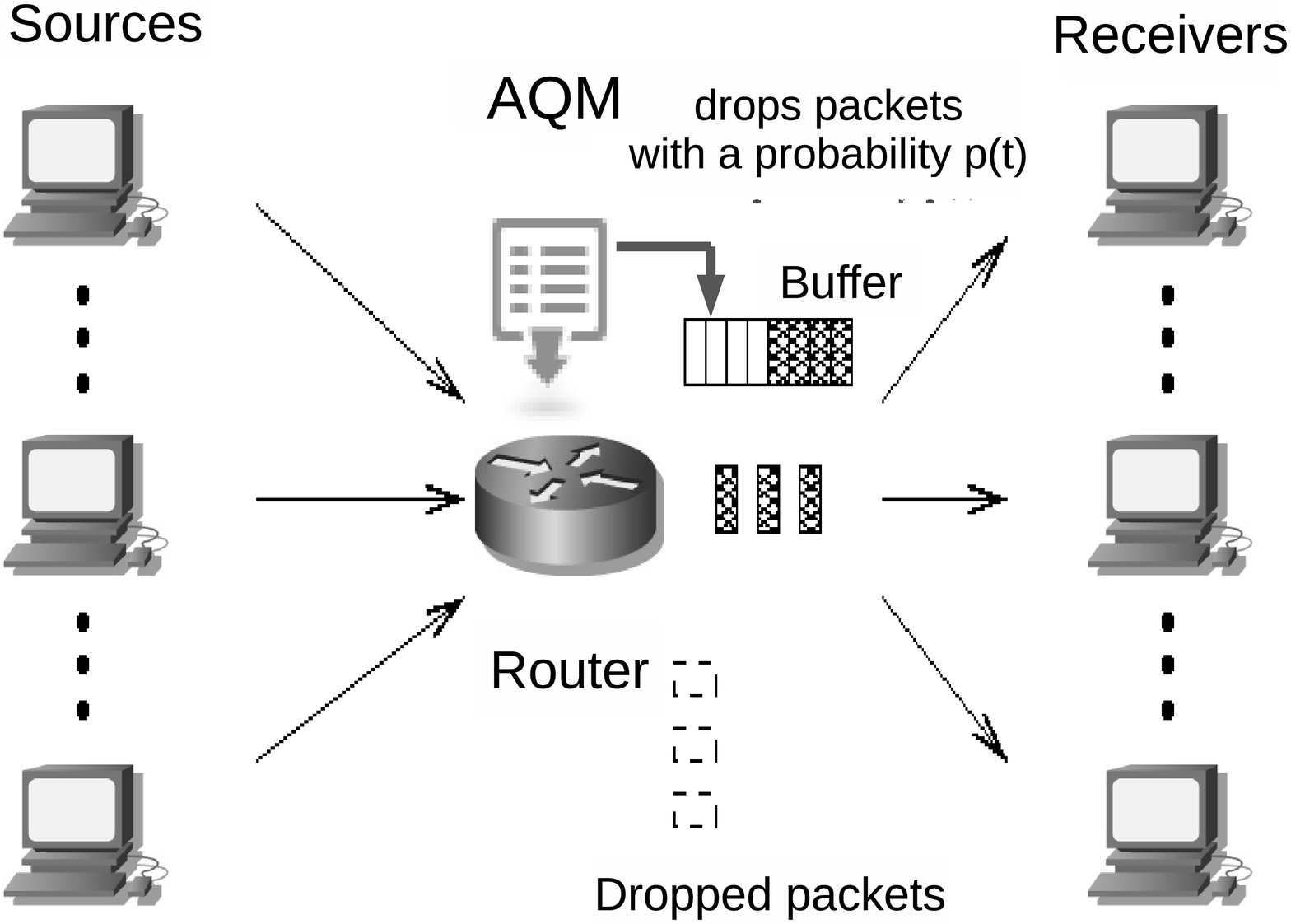}}
       \caption{Network control}
       \label{topologie2}
\end{figure}

Deterministic fluid-flow models have been widely used (see \cite{Low02}, \cite{Mis00}, \cite{Kel98} and \cite{Sri04} and references therein) to describe congestion control and AQM schemes in IP networks. These models capture the mean behavior of the TCP dynamic. While many studies dealing with network control in the automatic control theory framework consider the model proposed by \cite{Mis00}, we use in this paper the model introduced in \cite{Low02} and described by  (\ref{modelNL}). Contrary to the former, the model (\ref{modelNL}) takes into account the forward and backward delays and does not make the simplifying assumption that $(W(t-\tau)/W(t))(1-p(t-\tau^b))= 1$ \cite{Kim06}. The model and notations are as follow:
\begin{equation}
  \label{modelNL}
\left\{\begin{array}{rl}
\dot{W}_i(t)&=\frac{W_i(t-\tau_i)}{\tau_i(t-\tau_i)}(1-p_i(t-\tau_i^b))\frac{1}{W_i(t)}\\&~~~~-\frac{W_i(t-\tau_i)}{\tau_i(t-\tau_i)}\frac{W_i}{2}p_i(t-\tau_i^b)\\
\dot{b}(t)&=-C+\sum_N\eta_i\frac{W_i(t-\tau_i^f)}{\tau_i(t-\tau_i^f)}\\
\tau_i&=\frac{b(t)}{C}+T_{p_i}=\tau_i^f+\tau_i^b
\end{array} \right.
\end{equation}
where  $W_i(t)$ is the congestion window size of the source $i$, $b(t)$ is the queue length of the buffer at the router, $\tau_i$ is the round trip time (RTT) perceived by the source $i$. This latter quantity can be decomposed as the sum of the forward and backward delays ($\tau_i^f$ and $\tau_i^b$), standing for, respectively, the trip time from the source $i$ to the router (the one way) and from the router to the source via the receiver (the return) (see Figure \ref{topologie4}). $C$, $T_{p_i}$ and $N$ are parameters related to the network configuration and represent, respectively, the link capacity, the propagation time of the path taken by the connection $i$ and the number of TCP sources. $\eta_i$ is the number of sessions established by source $i$. The signal $p_i(t)$ corresponds to the drop probability of a packet.\\~\indent
In this paper, the objective is to develop a method which computes the appropriate dropping probability applied at the router in order to regulate the queue length of the buffer $b(t)$ to a desired level (Figure \ref{topologie2}). Since control depends on the system state, it is required to have access to them. However, congestion windows $W_i$ are not measurable. So that, we propose to reformulate the model (\ref{modelNL}) such that state vector can be measured. To this end, rates of each flow $x_i$, expressed as $x_i(t)=\frac{W_i(t)}{\tau_i(t)}$, will be considered. Hence, the dynamic of this new quantity is of the form $\dot{x_i}(t)=\frac{d}{dt}\left(\frac{W_i(t)}{\tau_i(t)}\right)=\frac{\dot{W_i}(t)-x_i(t)\dot{\tau_i}(t)}{\tau_i(t)}$. Based on the expressions of $\dot{W}(t)$, $\dot{b}(t)$, $\tau_i(t)$ (see equation (\ref{modelNL})) and $\dot{\tau}(t)=\frac{\dot{b}(t)}{C}$, a new model of the TCP behavior is derived

\begin{equation}
\label{modelNL2}
\left\{\begin{array}{rl}
\dot{x}(t)&=\frac{x(t-\tau)}{x(t)\tau(t)^2}(1-p(t-\tau ^b))-\frac{x(t-\tau)x(t)}{2}p(t-\tau^b)\\&~~~~+\frac{x(t)}{\tau(t)}-\frac{x(t)}{\tau(t)C}\sum_N\eta_ix_i(t-\tau^f_i)\\
\dot{b}(t)&=-C+\sum_N\eta_ix_i(t-\tau_i^f)
\end{array} \right..
\end{equation}
\begin{remark}
This model transformation allows us to use $x_i$ instead of $W_i$ which is more suitable to handle. Indeed, numerous works have developed tools that enable flow rates measurements, especially in anomaly detection framework (see for example \cite{Bar01}, \cite{Kim05}). Besides, The measure of the aggregate flow has already been proposed and successfully exploited in \cite{Kun01} and \cite{Kim06} for the realization of the AVQ ({\it Adaptive Virtual Queue}) and a PID type AQM respectively.
\end{remark}
~\indent Our work focuses on the congestion control of a single router with a static topology ($N$ and $\eta_i$ are constant). Moreover, for the mathematical tractability, we make the usual assumption \cite{Low02}, \cite{Hol02}, \cite{Kim06} that all delays ($\tau_i$, $\tau_i^f$ and $\tau_i^b$) are time invariant when they appear as argument of a variable (for example $x_i(t-\tau_i(t))\equiv x_i(t-\tau_i)$). This latter assumption is valid as long as the queue length remains close to its equilibrium value and when the queueing delay is smaller than propagation delays. Defining an equilibrium point
\begin{equation}
  \label{point_eq}
\left\{\begin{array}{rl}
\tau_{i_0}&=T_p+b_0/C\\
\dot{b}(t)&=0~\Rightarrow~\sum_N\eta_ix_{i_0}=C\\
\dot{x}_i(t)&=0~\Rightarrow~p_{i_0}=\frac{2}{2+(x_{i_0}\tau_{i_0})^2}
\end{array} \right.,
\end{equation}
model (\ref{modelNL2}) can be linearized:
{\small
\begin{equation}
\label{forme_can}
\begin{aligned}
\left[\begin{array}{c}
\dot{x}_1(t)\\\vdots\\\dot{x}_N(t)\\\dot{b}(t)\end{array}\right]=&
A\left[\begin{array}{c}
\delta x_1(t)\\\vdots\\ \delta x_N(t)\\\delta b(t)\end{array}\right]+
A_d
\left[\begin{array}{c}
\delta x_1(t-\tau_1^f)\\\vdots\\\delta x_N(t-\tau_N^f)\\\delta b(t)\end{array}\right]\\&~~+
B\left[\begin{array}{c}
\delta p_1(t-\tau_1^b)\\ \vdots\\ \delta p_N(t-\tau_N^b)\end{array}\right]
\end{aligned}
\end{equation}}
where $\delta x_i \doteq x_i-x_{i_0}$, $\delta b \doteq b-b_0$ and $\delta p_i
\doteq p_i-p_{i_0}$ are the state variations around the equilibrium point (\ref{point_eq}). Matrices of the equation (\ref{forme_can}) are defined by
\begin{equation*}
\begin{aligned}
A=&\left[\begin{array}{cccc}a_1&\sf{0}&\sf{0}&h_1\\\sf{0}&\ddots&\sf{0}&\vdots\\\sf{0}&\sf{0}&a_N&h_N\\\sf{0}&\sf{0}&\sf{0}&\sf{0}\end{array}\right],~~B=\left[\begin{array}{ccc}e_1&0&0\\0&\ddots&0\\0&0&e_N\\0&0&0\end{array}\right]\\
A_d=&\left[\begin{array}{cccc}f_1\eta_1&\ldots&f_1\eta_N&\sf{0}\\\vdots&\vdots&\vdots&\sf{0}\\f_N\eta_1&\ldots&f_N\eta_N&\sf{0}\\\eta_1&\ldots&\eta_N&\sf{0}\end{array}\right],
\end{aligned}
\end{equation*}
with $a_i=-\frac{1-p_{i_0}}{x_{i_0}\tau^2_{i_0}}-\frac{x_{i_0}p_{i_0}}{2}$, $h_i=-\frac{2(1-p_{i_0})}{C\tau^3_{i_0}}$, $f_i=-\frac{x_{i_0}}{\tau_{i_0}C}$ and $e_i=-\frac{1}{\tau_{i_0}^2}-\frac{x_{i_0}^2}{2}$.
Remark that a multiple time delays system \eqref{forme_can} is obtained with a particular form since each component of the state vector is delayed by a different quantity related to the communication path. Thus, in Section \ref{synthesis} an appropriate modeling of \eqref{forme_can} with a structured form (in quadratic separation framework) is proposed in order to formulate the stability condition.\\
\begin{figure}
       \centerline{\includegraphics[height=.25 \textheight, width=8cm]{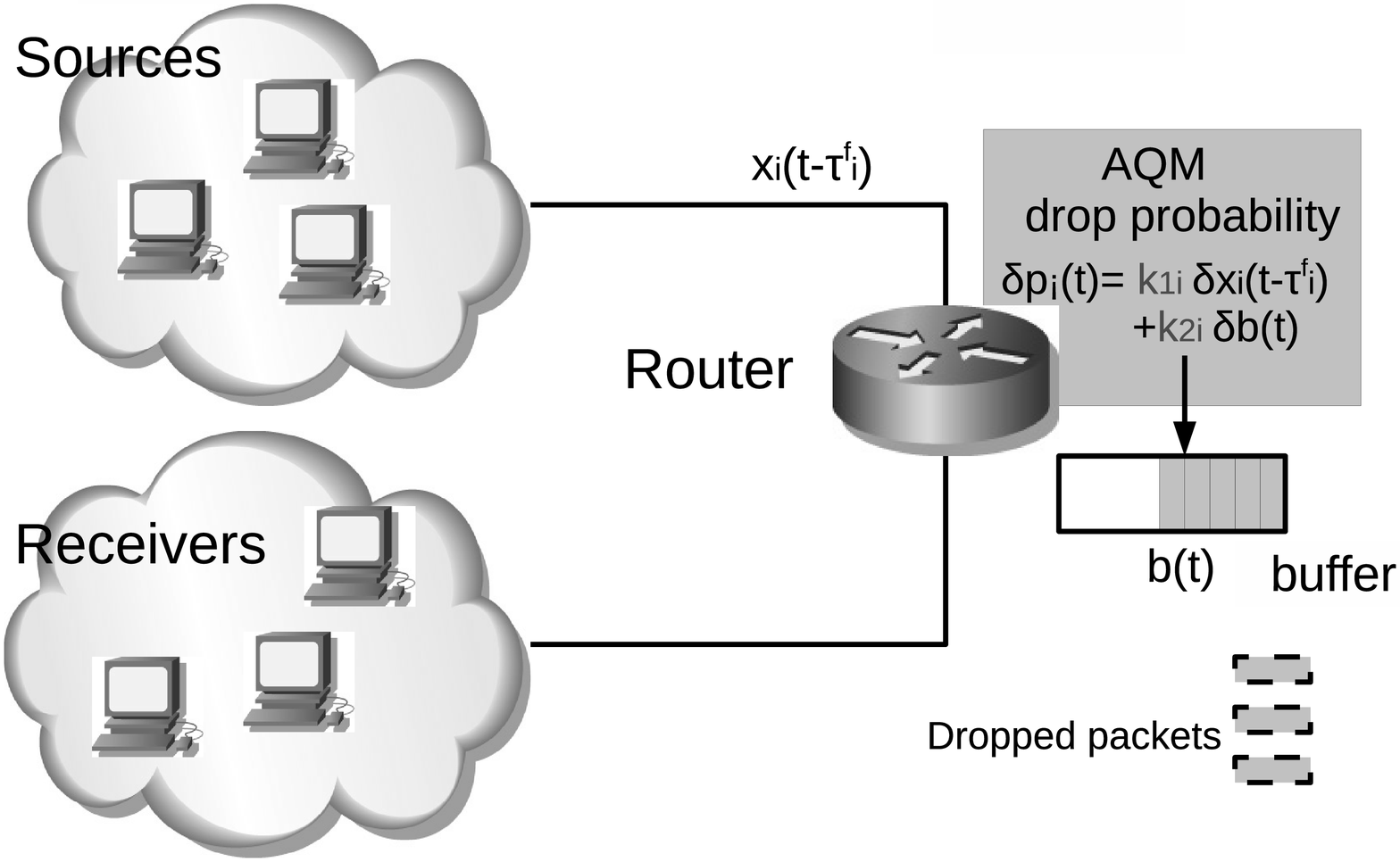}}
       \caption{Implementation of an AQM}
       \label{schema_block_struct}
\end{figure}
~\indent The problem of regulation is tackled in Section \ref{synthesis} with the design of a stabilizing state feedback for multiple time delays systems in order to guarantee a QoS. Hence, the dropping probability $p_i$ associated to source $i$ will be computed at the router by
\begin{equation}
\label{statefeedback}
p_i(t)=p_{i_0}+k_1\delta x_i(t-\tau_i^f)+k_2\delta b(t)
\end{equation}
with $k_1$, $k_2$ are components which have to be designed (see Figure \ref{schema_block_struct}). Note that the dropping probability $p_i$ perceived by the source $i$ will be delayed $p_i(t-\tau_i^b)$ because of the backward delay. Thus, a structured state feedback of the form \eqref{statefeedback} is proposed:
\begin{itemize}
\item to avoid unnatural signals with different delay combinations in the state feedback: $\delta x_i(t-\tau_i^f-\tau_j^b)$ for $i,~j\in\{1,...,N\}$ introducing then many additional delays,
\item it provides light computations (with less operations) reducing then the processing time at router,
\item it provides a decentralized control if the dropping strategy is performed at end hosts when emulating AQM \cite{Bha07}.
\end{itemize}
 ~\indent Applying the structured state feedback type control law $p_i(t)$  to each source $i$, $\forall i\in\{1,\ldots,N\}$,  the following interconnected system is obtained
{\small
\begin{equation}
\begin{aligned}
\label{interconnexion}
\left[\!\begin{array}{c}\delta\dot{x}_1(t)\\\vdots\\\delta \dot{x}_N(t)\\ \dot{b}(t)\end{array}\!\right]=&A\left[\!\begin{array}{c}\delta x_1(t)\\\vdots\\\delta x_N(t)\\ \delta b(t)\end{array}\!\right]+
A_d\left[\!\begin{array}{c}\delta x_1(t-\tau_1^f)\\\vdots\\\delta x_N(t-\tau_N^f)\\ \delta b(t)\end{array}\!\right]\\+&BK_1\left[\!\begin{array}{c}\delta x_1(t-\tau_1)\\ \vdots\\ \delta x_N(t-\tau_N)\end{array}\!\right]+BK_2\left[\!\begin{array}{c}\delta b(t-\tau_1^b)\\ \vdots\\ \delta b(t-\tau_N^b)\end{array}\!\right]
\end{aligned}
\end{equation}
}
where matrices gains $K_1$ and $K_2$ are structured as $K_1=diag\{k_{11},\ldots,k_{1
N}\}$ and $K_2=diag\{k_{21},\ldots,k_{2N}\}$.

Equation (\ref{interconnexion}) represents thus the mean behavior of TCP regulated by a structured state feedback type AQM around an equilibrium point. $K_1$ and $K_2$ can be derived from the stability analysis of the interconnection
(\ref{interconnexion}). To this end, we propose to design these latter gains by a suitable modeling of (\ref{interconnexion}) applying then the quadratic separation principle \cite{Pea07}, \cite{Gou06a} for the stability condition.

%#########################
\section{STABILIZATION AND QoS GUARANTEE: DESIGN OF AN AQM}
\label{synthesis}

The stability of the interconnected system (\ref{interconnexion}) depends on matrices $K_1$ and $K_2$. This section aims to develop a method that provides such stabilizing matrices. Thus, the quadratic separation framework is considered and the following theorem will be employed \cite{Pea07}.
\begin{figure}
       \centerline{\includegraphics[height=.06 \textheight]{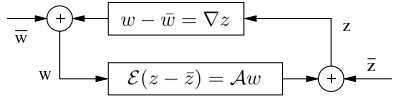}}
       \caption{An interconnected system}
       \label{feedbacksystem}
\end{figure}
\begin{theorem}
\label{theorem}
Given two possibly non-squared matrices $\mathcal{E}$, $\mathcal{A}$ and an uncertain matrix
 $\nabla$ belonging to a set $\Xi$. The uncertain system represented on Figure \ref{feedbacksystem} is stable for all matrices $\nabla\in\Xi$ if and only if it exists a matrix
 $\Theta=\Theta^*$ satisfying conditions
\begin{align}
\label{cond-theorem1}
\left[\begin{array}{cc}\mathcal{E}&-\mathcal{A}\end{array}\right]^{\perp*}\Theta
\left[\begin{array}{cc}\mathcal{E}&-\mathcal{A}\end{array}\right]^{\perp}>\sf{0}\\
\label{cond-theorem2}
\left[\begin{array}{cc}\sf{1}&\nabla^*\end{array}\right]\Theta
\left[\begin{array}{c}\sf{1}\\ \nabla\end{array}\right]\leq\sf{0}.
\end{align}
\end{theorem}

It is then required to transform the initial system (\ref{interconnexion}) into a feedback system of the form of  Figure \ref{feedbacksystem} where a linear equation is connected to a linear uncertainty $\nabla$. The key idea (proposed by \cite{Gou06a} for the single time-delay systems case) consists in associating the delay operator as an uncertainty which must be bounded. Hence, Theorem \ref{theorem} may be applied to the multiple time delays system (\ref{interconnexion}) by rewritting it as an interconnected system (see Figure \ref{feedbacksystem}) with $\mathcal{E}=\sf{1}$,
\begin{equation}
\label{relation-nabla-0}
\overbrace{\left[\begin{array}{c} X(t)\\
\hat{x}^{\tau^f}(t)\\
\hat{b}^{\tau^b}(t)\\
\hat{x}^{\tau}(t)
\end{array}\right]}^{w}=
\overbrace{\left[\begin{array}{cccc} s^{-1}\sf{1}_4&&&\\ &\mathcal{D}_{\tau^f}&&\\  &&\mathcal{D}_{\tau^b}&\\&&&\mathcal{D}_{\tau}
\end{array}\right]}^{\nabla}
\overbrace{
\left[\begin{array}{c} \dot{X}(t)\\
\hat{x}(t)\\
\hat{b}(t)\\
\hat{x}(t)
\end{array}\right]}^{z}
\end{equation}
and
\begin{equation}
\label{relation-EA-0}
\mathcal{E}\overbrace{\left[\begin{array}{c}\dot{X}(t)\\
\hat{x}(t)\\
\hat{b}(t)\\
\hat{x}(t)\end{array}\right]}^{z}=\overbrace{\left[\begin{array}{cccc}A&\bar{A}_d&BK_2&BK_1\\
E_1&\sf{0}&\sf{0}&\sf{0}\\
E_2&\sf{0}&\sf{0}&\sf{0}\\
E_1&\sf{0}&\sf{0}&\sf{0}
\end{array}\right]}^{\mathcal{A}}\overbrace{\left[\begin{array}{c} X(t)\\
\hat{x}^{\tau^f}(t)\\
\hat{b}^{\tau^b}(t)\\
\hat{x}^{\tau}(t)\end{array}\right]}^{w}
\end{equation}
where
\begin{equation}
\begin{aligned}
\label{def-signals}
\hat{b}(t)=&\left[\begin{array}{c}b(t)\\ \vdots\\b(t) \end{array}\right],\hat{b}^{\tau^b}(t)=\left[\begin{array}{c}b(t-\tau_1^b)\\ \vdots\\b(t-\tau_N^b) \end{array}\right],\\ \hat{x}^{\tau^f}(t)=&\left[\begin{array}{c}x_1(t-\tau_1^f)\\ \vdots\\x_N(t-\tau_N^f) \end{array}\right],\hat{x}^{\tau}(t)=\left[\begin{array}{c}x_1(t-\tau_1)\\ \vdots\\x_N(t-\tau_N) \end{array}\right],\\
\hat{x}(t)=&\left[\begin{array}{c}x_1(t)\\ \vdots\\x_N(t) \end{array}\right],\mathcal{D}_\Diamond=\left[\begin{array}{ccc}e^{-\Diamond_1s}&&\sf{0}\\&\ddots&\\\sf{0}&&e^{-\Diamond_Ns}\end{array}\right],
\end{aligned}
\end{equation}
$X(t)=[\hat{x}'~~b(t)]'$. The delay matrix operators $\mathcal{D}_\Diamond$ ($\Diamond$ represents $\tau^f$, $\tau^b$ or $\tau$) must be defined to create the delayed signals $\hat{x}^{\tau^f}$, $\hat{x}^{\tau}$ and $\hat{b}^{\tau^b}$ (\ref{def-signals}).\\~\indent
We aim at proving the stability ({\it i.e.} no poles in the right hand side of the complex plane for all values of the delay and for all values of the uncertainty $\nabla\in\Xi$) which problem can be recast in the present framework as the well-posedness of the feedback system for all $s\in\sf{C}^+$, for all values of the delays ($\tau_i$, $\tau_i^{f}$ and $\tau_i^{b}$, $i=\{1,\ldots,N\}$) and all admissible uncertainties $\nabla\in\Xi$. Then a conservative choice of quadratic separator that fullfils (\ref{cond-theorem2}) is of the form
\vspace*{-0.1cm}
\begin{equation}
\label{separateur0}
\Theta=\left[\begin{array}{cccc|cccc}\sf{0}&\sf{0}&\sf{0}&\sf{0}&-P&\sf{0}&\sf{0}&\sf{0}\\
\sf{0}&-Q^f&\sf{0}&\sf{0}&\sf{0}&\sf{0}&\sf{0}&\sf{0}\\
\sf{0}&\sf{0}&-Q^b&\sf{0}&\sf{0}&\sf{0}&\sf{0}&\sf{0}\\
\sf{0}&\sf{0}&\sf{0}&-Q&\sf{0}&\sf{0}&\sf{0}&\sf{0}\\ \hline
-P&\sf{0}&\sf{0}&\sf{0}&\sf{0}&\sf{0}&\sf{0}&\sf{0}\\
\sf{0}&\sf{0}&\sf{0}&\sf{0}&\sf{0}&Q^f&\sf{0}&\sf{0}\\
\sf{0}&\sf{0}&\sf{0}&\sf{0}&\sf{0}&\sf{0}&Q^b&\sf{0}\\
\sf{0}&\sf{0}&\sf{0}&\sf{0}&\sf{0}&\sf{0}&\sf{0}&Q
\end{array}\right],
\end{equation}
with $P\in\sf{R}^{N+1\times N+1}>\sf{0}$ and $Q^\Diamond=diag(q^\Diamond_1,\ldots,q^\Diamond_N)$ where $q^\Diamond_i$ are positive scalars for all $i=\{1,\ldots,N\}$ and for $\Diamond=\{f,b,\emptyset\}$ (see \cite{Gou06a} for a simpler case).\\
~\indent Then, it remains to test the first condition (\ref{cond-theorem1}). Since this inequality does not depend on delays, the derived condition is said to be {\it Independent Of Delays} (IOD). Consequently, this latter criterion provides state feedback gains that stabilize the system for all possible values of delays. It thus appears that this method is very conservative and it would be interesting to have a condition depending on delays.\\
~\indent We aim now at deriving a {\it Delay Dependent} result ( $i.e.$ the well known DD approach which ensures the stability for all values of the delay between zero and an upper bound) with the same methodology. To do so note that the results were delay independent because operators $e^{-\Diamond_is}$, when $s\in\sf{C}^+$, can only be characterized as uncertainties norm bounded by $1$. To get delay dependent results it is therefore needed to have characteristics that depend on upper bounds of delays. This can be done noting that for all $s\in\sf{C}^+$ and a given delay $h\in[0 ~~\bar{h}]$ one has $|s^{-1}(1-e^{-hs})|\leq\bar{h}$ and this operator is such that $V(s)=s^{-1}(1-e^{-hs})\dot{X}(s)$ where $V(s)$ and $\dot{X}(s)$ are the Laplace transforms respectively of $v(t)=x(t)-x(t-h)$ and $\dot{x}(t)$. Introducing this new operator for each delay $\tau_i$, $\tau_i^{f}$ and $\tau_i^{b}$, $i=\{1,\ldots,N\}$ leads to write the delay dependent stability problem of (\ref{interconnexion}) as a well-posedness problem of the system in Figure \ref{feedbacksystem} with
{\small
\begin{equation}
\label{relationEA}
\begin{aligned}
\!\!\! \mathcal{E}&=\left[\begin{array}{ccccccc}
\!\!\!\sf{1}&\sf{0}&\sf{0}&\sf{0}&\sf{0}&\sf{0}&\sf{0}\!\!\!\\\!\!\!\sf{0}&\sf{1}&\sf{0}&\sf{0}&\sf{0}&\sf{0}&\sf{0}\!\!\!\\\!\!\!&\sf{0}&\sf{1}&\sf{0}&\sf{0}&\sf{0}&\sf{0}\!\!\!\\\!\!\!\sf{0}&\sf{0}&\sf{0}&\sf{1}&\sf{0}&\sf{0}&\sf{0}\!\!\!\\\!\!\!E_1&\sf{0}&\sf{0}&\sf{0}&-\sf{1}&\sf{0}&\sf{0}\!\!\!\\\!\!\!E_2&\sf{0}&\sf{0}&\sf{0}&\sf{0}&-\sf{1}&\sf{0}\!\!\!\\\!\!\!E_1&\sf{0}&\sf{0}&\sf{0}&\sf{0}&\sf{0}&-\sf{1}\!\!\!\\\!\!\!
\sf{0}&\sf{0}&\sf{0}&\sf{0}&\sf{0}&\sf{0}&\sf{0}\!\!\!\\\!\!\!\sf{0}&\sf{0}&\sf{0}&\sf{0}&\sf{0}&\sf{0}&\sf{0}\!\!\!\\\!\!\!\sf{0}&\sf{0}&\sf{0}&\sf{0}&\sf{0}&\sf{0}&\sf{0}
\!\!\!\end{array}\right],w=\left[\begin{array}{c}\!\!\! X(t)\!\!\!\\\!\!\!
\hat{x}^{\tau^f}(t)\!\!\!\\\!\!\!
\hat{b}^{\tau^b}(t)\!\!\!\\\!\!\!
\hat{x}^{\tau}(t)\!\!\!\\\!\!\!
w_1(t)\!\!\!\\\!\!\!
w_2(t)\!\!\!\\\!\!\!
w_3(t)\!\!\!\\\!\!\!
\end{array}\right],\\
\mathcal{A}&=\left[\begin{array}{ccccccc}
\!\!\!A&\bar{A}_d&BK_2&BK_1&\sf{0}&\sf{0}&\sf{0}\!\!\!\\\!\!\!E_1&\sf{0}&\sf{0}&\sf{0}&\sf{0}&\sf{0}&\sf{0}\!\!\!\\\!\!\!E_2&\sf{0}&\sf{0}&\sf{0}&\sf{0}&\sf{0}&\sf{0}\!\!\!\\\!\!\!E_1&\sf{0}&\sf{0}&\sf{0}&\sf{0}&\sf{0}&\sf{0}\!\!\!\\\!\!\!\sf{0}&\sf{0}&\sf{0}&\sf{0}&\sf{0}&\sf{0}&\sf{0}\!\!\!\\\!\!\!\sf{0}&\sf{0}&\sf{0}&\sf{0}&\sf{0}&\sf{0}&\sf{0}\!\!\!\\\!\!\!\sf{0}&\sf{0}&\sf{0}&\sf{0}&\sf{0}&\sf{0}&\sf{0}\!\!\!\\\!\!\!
E_1&-\sf{1}&\sf{0}&\sf{0}&-\sf{1}&\sf{0}&\sf{0}\!\!\!\\\!\!\!E_2&\sf{0}&-\sf{1}&\sf{0}&\sf{0}&-\sf{1}&\sf{0}\!\!\!\\\!\!\!E_1&\sf{0}&\sf{0}&-\sf{1}&\sf{0}&\sf{0}&-\sf{1}\!\!\!
\end{array}\right],z=
\left[\begin{array}{c}\!\!\! \dot{X}(t)\!\!\!\\ \!\!\!
\hat{x}(t)\!\!\!\\\!\!\!
\hat{b}(t)\!\!\!\\\!\!\!
\hat{x}(t)\!\!\!\\\!\!\!
\dot{\hat{x}}(t)\!\!\!\\\!\!\!
\dot{\hat{b}}(t)\!\!\!\\\!\!\!
\dot{\hat{x}}(t)\!\!\!\\\!\!\!
\end{array}\right],
\end{aligned}
\end{equation}}
where {\small $\bar{A}_d=A_d\left[\begin{array}{c}\sf{1}_N\\\sf{0}_{1\times N}\end{array}\right],~E_1=\left[\begin{array}{cc}\sf{1}_N&\sf{0}_{N\times 1}\end{array}\right],~E_2=\left[\begin{array}{c|c}\sf{0}_N&\begin{array}{c}1\\\vdots\\1\end{array}\end{array}\right]$}
and $
w_1(t)=\hat{x}(t)-\hat{x}^{\tau^f}(t),~w_2(t)=\hat{b}(t)-\hat{b}^{\tau^b}(t),~w_3(t)=\hat{x}(t)-\hat{x}^{\tau}(t)$ and with the augmented uncertain operator
{\small \begin{equation}
\label{relation-nabla}
%\overbrace{\left[\begin{array}{c} X(t)\\
%\hat{x}^{\tau^f}(t)\\
%\hat{b}^{\tau^b}(t)\\
%\hat{x}^{\tau}(t)\\
%\hat{x}(t)-\hat{x}^{\tau^f}(t)\\
%\hat{b}(t)-\hat{b}^{\tau^b}(t)\\
%\hat{x}(t)-\hat{x}^{\tau}(t)\\
%\end{array}\right]}^{w}=
\nabla=Diag\left( s^{-1}\sf{1}_4, \mathcal{D}_{\tau^f},  \mathcal{D}_{\tau^b},
\mathcal{D}_{\tau},\mathcal{I}_{\tau^f},\mathcal{I}_{\tau^b},\mathcal{I}_{\tau}\right)
%\overbrace{
%\left[\begin{array}{c} \dot{X}(t)\\
%\hat{x}(t)\\
%\hat{b}(t)\\
%\hat{x}(t)\\
%\dot{\hat{x}}(t)\\
%\dot{\hat{b}}(t)\\
%\dot{\hat{x}}(t)\\
%\end{array}\right]}^{z}
\end{equation}}
where $\mathcal{I}_\Diamond=diag(\frac{1-e^{-\Diamond_1s}}{s},\ldots,\frac{1-e^{-\Diamond_Ns}}{s})$, $\Diamond=\{f,b,\emptyset\}$. Delay operators $\mathcal{D}_\Diamond$ are thus isolated as well as operators $\mathcal{I}_\Diamond$ ensuring a better bound on each delays ({\it delay dependent} case, see \cite{Gou06a}). Expressing the covering set on every block of $\nabla$, a conservative choice of separator satisfying the second inequality (\ref{cond-theorem2}) of Theorem \ref{theorem} is
\begin{equation}
\label{separator1}
\Theta=\left[\begin{array}{cc}
\Theta_{11}&\Theta_{12}\\\Theta_{12}'&\Theta_{22}
\end{array}\right]
\end{equation}
with
\begin{equation}
\begin{aligned}
\label{def-theta-separator}
\Theta_{11}=&diag\!\left(\sf{0}_{N+1},\!-Q_0^f,\!-Q_0^b,\!-Q_0,\!-Q_1^fT^{f^2}\!\!,\!-Q_1^bT^{b^2}\!\!,\!-Q_1T^2\!\right)\!,\\
\Theta_{12}=&diag\!\left(-P,\sf{0}_{6N}\right),\\
\Theta_{22}=&diag\!\left(\sf{0}_{N+1},Q_0^f,Q_0^b,Q_0,Q_1^f,Q_1^b,Q_1\right)\!,\\
Q_0^\Diamond=&diag\!\left(q_{01}^\Diamond,\ldots,q_{0N}^\Diamond\right)\!,\\
Q_1^\Diamond=&diag\!\left(q_{11}^\Diamond,\ldots,q_{1N}^\Diamond\right)\!,\\
T^\Diamond=&diag\!\left(\tau_1^\Diamond,\ldots,\tau_N^\Diamond\right)\!,
\end{aligned}
\end{equation}
where $P\in\sf{R}^{N+1\times N+1}>\sf{0}$ and $q^\Diamond_i$ are positive scalars for all $i=\{1,\ldots,N\}$ and for $\Diamond=\{f,b,\emptyset\}$. Since, the inequality (\ref{cond-theorem2}) is satisfied by definition of the operator bounds of the uncertain matrix $\nabla$, it remains to verify the first one (\ref{cond-theorem1}). After some algebraic manipulations, it can be shown that this latter inequality can be expressed as
\begin{equation}
\label{condition-finale}
\left[\begin{array}{ccc}
\Xi_{1}&\Xi_2Q_1T&\Xi_2Q_1^fT^f\\
\ast&Q_1T&\sf{0}_{N}\\
\ast&\ast&Q_1^fT^f
\end{array}\right]>\sf{0}
\end{equation}
where
\begin{equation}
\begin{aligned}
\Xi_{1}=&N_1^T\Theta N_1+N_2^T\Theta N_1+N_1^T\Theta N_2,\\
\Xi_{2}=&\left[\begin{array}{ccc}\sf{0}_{N\times 2N+1}&E_1BK_2&E_1BK_1\end{array}\right]^T,\\
N_1=&\left[\begin{array}{c}
\begin{array}{ccc}A&\bar{A}_d&\sf{0}_{N+1\times 2N}\end{array}\\ \hline
\begin{array}{cc}\begin{array}{c}E_1\\ E_2\\ E_1\end{array}&\sf{0}_{3N\times 3N}\end{array}\\ \hline
\begin{array}{ccc}E_1A&E_1\bar{A}_d&\sf{0}_{N\times 2N}\\
\sf{0}_{N\times N+1}&E_2\bar{A}_d&\sf{0}_{N\times 2N}\\
E_1A&E_1\bar{A}_d&\sf{0}_{N\times 2N}\end{array}\\ \hline
\sf{1}_{4N+1}\\ \hline
\begin{array}{cc}\begin{array}{c}E_1\\ E_2\\ E_1\end{array}&-\sf{1}_{3N}\end{array}\\
\end{array}\right],\\
N_2=&\left[\begin{array}{c}
\begin{array}{ccc}\sf{0}_{N+1\times 2N+1}&BK_2&BK_1\end{array}\\ \hline
\sf{0}_{3N\times 4N+1}\\ \hline
\begin{array}{ccc}\sf{0}_{N\times 2N+1}&E_1BK_2&E_1BK_1\end{array}\\ \hline 
\sf{0}_{N\times 4N+1}\\ \hline
\begin{array}{ccc}\sf{0}_{N\times 2N+1}&E_1BK_2&E_1BK_1\end{array}\\ \hline
\sf{0}_{7N+1\times 4N+1}\\
\end{array}\right].
\end{aligned}
\end{equation}
 This latter condition gives the following theorem:
\begin{theorem}
\label{theorem2}
For given scalars $\tau_i$, $\tau_i^b$ and $\tau_i^f$ for $i=\{1,...,N\}$, if there exists a $N+1\times N+1$ positive definite matrix $P$ and $N\times N$ diagonal positive matrices $Q_k^f$, $Q_k^b$, $Q_k$ with $k=\{0,1\}$ and $K_1$, $K_2$ such that the inequality  (\ref{condition-finale}) is satisfied, then the system (\ref{interconnexion}) is stable.
\end{theorem}
The state feedback gains  $K_1$ and $K_2$ are thus derived solving the inequality \eqref{condition-finale} of Theorem \ref{theorem2}. Regarding the synthesis problem, since $K_1$ and $K_2$ are decision variables, condition (\ref{condition-finale}) is bilinear and a global optimal solution cannot be found. Nevertheless, the feasibility problem can still be tested to provide a suboptimal solution:
\begin{enumerate}
\item[$\vartriangleright$] using a BMI solver as penbmi \cite{penbmi}.
\item[$\vartriangleright$] using a relaxation algorithm \cite{Nic01}:
\begin{itemize}
\item $step~0$. Initialization: $K_1=K_{1_0}$, $K_2=K_{2_0}$.
\item $step~1$. Run LMI computation (inequality \eqref{condition-finale} with fixed $K_1$, $K_2$) $\Rightarrow$ backup $P_{0}=P$, $Q_{1_0}=Q_1$, $Q_{1_0}^f=Q_1^f$.
\item $step~2$. Set $P=P_{0}$, $Q_1=Q_{1_0}$, $Q_1^f=Q_{1_0}^f$ and $K_1$, $K_2$ are free, run LMI computation.
\item $step~3$. If condition feasible: stop algorithm. Otherwise: return to $step~0$
\end{itemize}
\end{enumerate}
 It is worthy to note that the state feedback gains $K_1$ and $K_2$ is easily and routinely derived solving the matrix inequality (\ref{condition-finale}) while many AQM such RED are well known to be difficult to tune as it has been stated in \cite{Chr00}, \cite{Bon00}.

\section{NS-2 SIMULATIONS}
\begin{figure}
       \centerline{\includegraphics[height=.21 \textheight]{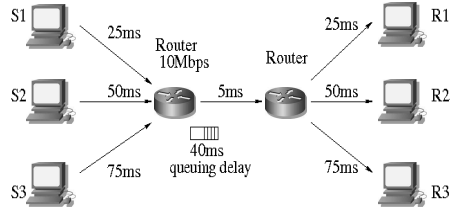}}
       \caption{Network topology: an example}
       \label{topologie_exemple}
\end{figure}

\begin{figure}
        \centerline{\includegraphics[height=13cm, width=9.5cm]{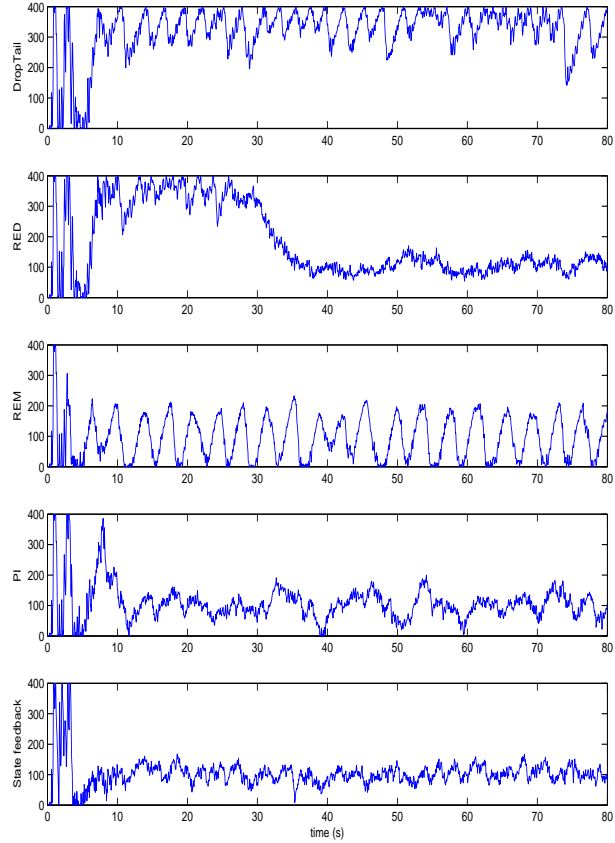}}
        \vspace*{-1.3cm}
       \caption{Evolution of the queue length $b(t)$ (pkt): the desired size is $100pkts$}
       \label{courbe1}
\end{figure}

\begin{figure}
        \centering
        \includegraphics[height=13cm, width=9.5cm]{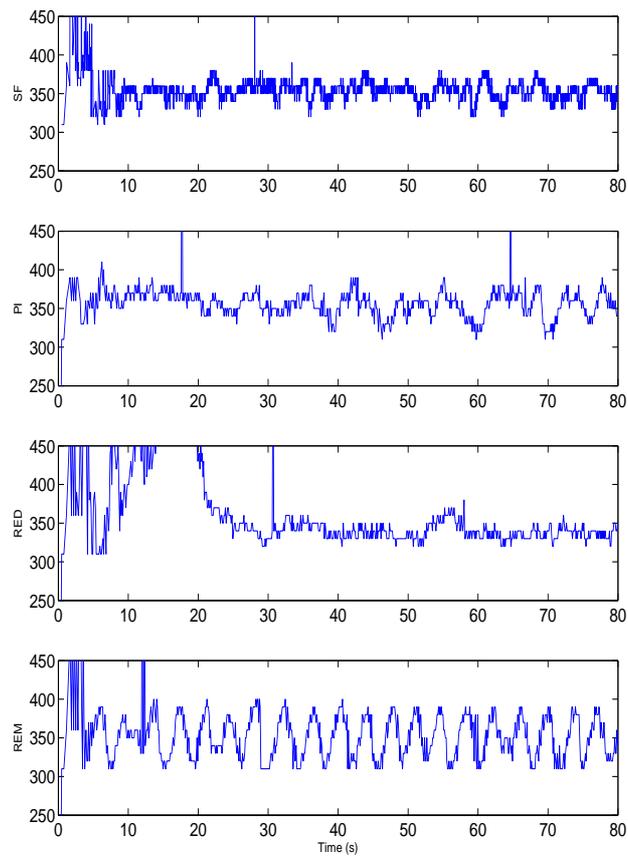}
       \caption{Evolution of the RTT of connections from source 3 (ms): the expected value is $350ms$}
       \label{rtt}
\end{figure}

In this section, we perform simulations with the network simulator NS-2 \cite{Fal02} (release 2.30) to validate the exposed theory. Throughout this part, the efficiency of the proposed mechanism is evaluated and compared to few existing AQM. Consider the numerical example of the Figure \ref{topologie_exemple}. So, the objective is to regulate the queue length of the router to a desired level $b_0=100$ packets while the maximal buffer size is set to $400$ packets. Propagation times are as illustrated on Figure \ref{topologie_exemple}. The link bandwidth is fixed to $10Mbps$, that is $2500$ packet/s considering packet size of $500$ bytes. Hence, at the equilibrium the queueing delay is equal to $40ms$. Each of the three sources uses TCP/Reno and establishes $10$ connections generating long lived TCP flows (like FTP connections). Upon these latter specifications, the equilibrium point (\ref{point_eq}) is derived: $b_0=100pkt$, $\tau_{1_0}=150ms$, $\tau_{2_0}=250ms$, $\tau_{3_0}=350ms$, $x_{1_0}=x_{2_0}=x_{3_0}=83.33pkt/s$ (rate for each connections of the three sources) and $p_0=10^{-3}\left[9.508~3.444~1.760\right]'$.
\begin{remark}
\label{remark4}
It is worthy to note that the second equation in (\ref{point_eq}) enables to choose arbitrarily the rate assigned to each sources. This latter option allows us to make a {\it service differentiation} between senders. In the previous example, we choose to assign the same sending rate for all sources (and for each connections) ensuring then fairness.
\end{remark}
The interconnected system of the form (\ref{interconnexion}) modeling a such network configuration is then written as
{\small
\begin{equation}
\label{interconnexion_exemple}
\begin{aligned}
&\left[\begin{array}{c}\delta\dot{x}_1(t)\\\delta\dot{x}_2(t) \\\delta \dot{x}_3(t)\\ \dot{b}(t)\end{array}\right]=\left[\begin{array}{cccc}
-2.02\!&0\!&0\!&-0.22\\
0\!&-0.75\!&0\!&-0.05\\
0\!&0\!&-3.88\!&-0.01\\
0\!&0\!&0\!&0
\end{array}\right]\left[\begin{array}{c}\delta x_1(t)\\\delta x_2(t)\\\delta x_3(t)\\ \delta b(t)\end{array}\right]\\&~~~+
\left[\begin{array}{cccc}
-2.22&-2.22&-2.22&0\\
-1.33&-1.33&-1.33&0\\
-0.95&-0.95&-0.95&0\\
20&20&20&0
\end{array}\right]\left[\begin{array}{c}\delta x_1(t-0.025)\\\delta x_2(t-0.05)\\\delta x_3(t-0.075)\\ \delta b(t)\end{array}\right]\\&~~~+\left[\begin{array}{ccc}
-912&0&0\\
0&-884&0\\
0&0&-876\\
0&0&0
\end{array}\right]\left(K_1\left[\begin{array}{c}\delta x_1(t-0.15)\\ \delta x_2(t-0.25)\\ \delta x_3(t-0.35)\end{array}\right]\right.\\&~~~~~~~~~~~~~\left.+K_2\left[\begin{array}{c}\delta b(t-0.125)\\ \delta b(t-0.2)\\ \delta b(t-0.275)\end{array}\right]\right)
\end{aligned}
\end{equation}}
Then, we have to find matrices $K_1$ and $K_2$ such that the feedback system (\ref{interconnexion_exemple}) is stable and the regulation around the equilibrium point is ensured. Applying Theorem \ref{theorem2} exposed in section \ref{synthesis}, two stabilizing matrix gains 
{\small\begin{equation}
\label{exemple_gain}
K_1\!=\!10^{-3}\!\left[\begin{array}{ccc}
\!\!-0.09\!\!&0\!\!&0\!\!\!\\
\!\!0\!\!&0.61\!\!&0\!\!\!\\
\!\!0\!\!&0\!\!&0.76\!\!\!
\end{array}\right]\!,~K_2\!=\!10^{-3}\!\left[\begin{array}{ccc}
\!\!0.27\!\!&0\!\!&0\!\!\!\\
\!\!0\!\!&0.13\!\!&0\!\!\!\\
\!\!0\!\!&0\!\!&0.08\!\!\!
\end{array}\right]
\end{equation}}
 can be found. Considering the network of Figure \ref{topologie_exemple}, we have simulated the congestion phenomenon at a router applying the drop tail mechanism and different AQM. Effect of these latter dropping strategies on the queue length have been evaluated and are illustrated on Figure \ref{courbe1}. Table \ref{stats1} presents additional characteristics of the results. As expected, the drop tail mechanism maintains the queue size close to the buffer overflow (maximal buffer size: $400pkt$) involving many uncontrolled dropped packets. Futhermore, a such large queue size implies a large queueing delay and also large oscillations providing an important delay jitter (see Table \ref{stats1}).\\
~\indent In order to maintain a controlled (and desired) queueing delay with a low jitter, the queue length at the router should be regulated. This 
issue is tackled with the use of AQM (adjustments of the different setting parameters are shown in table \ref{tuning}).
 \begin{table}
  \centering
  \caption{Adjustment of parameter setting of each AQM}\label{tuning}
\begin{tabular}{|c|c|}
  \hline
  RED & $min_{th}$=50,$max_{th}$=300,$w_Q$=5.99e-06,$max_p$=0.1,$f_s$=160Hz  \\
  \hline
  REM & $\gamma$=0.003, $\Phi$=1.001,$q_{ref}$=100pkt \\
  \hline
  PI & a=1.483e-05,b=1.479e-05,$q_{ref}$=100pkt,$f_s$=160Hz   \\
  \hline
  SF & gains $K_1$ and $K_2$ (\ref{exemple_gain}), equilibrium point (\ref{point_eq}) \\
  \hline
\end{tabular}
\end{table}
On Figure \ref{courbe1}, control performances of AQM can be measured by two observations: the transient performance (in particular the speed of response) and the steady state error control. It can be observed that the state feedback ($SF$) (\ref{exemple_gain}) ensures an efficient control since the queue length matches its equilibrium faster than other cases. Secondly, $SF$ provides a better regulation causing less oscillations and good precision. These characteristics can be verified in Table \ref{stats1} which analyses each response from Figure \ref{courbe1}. In the simulation, AQM was designed to regulate queue size at $100pkt$ (equilibrium) avoiding severe congestion and ensuring then a stable queueing delay of $40ms$. First, in Table \ref{stats1}, the $mean$ (and $average~ queueing~ delay$ since the queueing delay is expressed by $b(t)/C$) shows the accuracy of each AQM according the desired prescribed queue length. Next, the $standard~ deviation$ (and $average~ delay ~jitter$) shows the ability of each AQM to maintain the queue length (and queueing delay) close to the desired equilibrium providing then an efficient regulation. Note that REM has the lowest queueing delay, however the objective is to guarantee the queue length stability inducing then a queueing delay equals to $40ms$. If a lower queueing delay is required, one just has to change the equilibrium point (reducing the desired queue length).\\~\indent
Besides, the stability of the congestion phenomenon keeping a stable queue length (and thus a stable queueing delay), allows to control the RTT ($\tau_i$ for $i=\{1,...,N\}$) for all sources to a desired value with low variations (see Figure \ref{rtt}). Thanks to the efficient regulation applied by $SF$, a QoS in terms of RTT and delay jitter is hence guaranteed. It provides thus an interesting feature which is relevant for streaming and real-time applications. Although RED performs a good regulation on the queue length and the RTT at the steady state, its response time is very slow which makes RED inefficient against short-lived traffic perturbations \cite{Ari08}.\\~\indent
 Table \ref{stats3} presents statistics related to the packet arrival rates of each user when different AQM are implemented at the router and bears out the accuracy and the control efficiency of the proposed SF. The prescribed equilibrium rate for each connection (according to (\ref{point_eq})) that establishes fairness is $x_{1_0}=x_{2_0}=x_{3_0}=83.33pkt/s$. As it can be seen, only $SF$ is able to keep arrival rates comparatively close to the equilibrium value. Futhermore, based on the Jain's fairness index $=\frac{( \sum x_i )^2 }{(n \sum x_i^2 )}$ \cite{Jai84} (the more the index is close to 1, the more the distribution of the resources is fair), we observe that $SF$ applies a fair strategy.

\begin{table}
  \centering
  \caption{Some statistics on the queue length for different AQM}\label{stats1}
\begin{tabular}{|c|c|c|c|c|c|}
  \hline
   & DT & RED & REM & PI & SF \\
  \hline
  $Mean (pkt)$& 317.6 & 103.8 & 94.8 & 99.4 & 102.3 \\
  \hline
  $Stand. dev. (pkt)$ & 84 & 21.7 & 70.7 & 35.1 & 23.3 \\
  \hline
 $ \begin{array}{c}Average\\queueing ~delay (ms)\end{array}$ & 127 & 41.5 & 37.9 & 39.8 & 40.9 \\
  \hline
  $\begin{array}{c}Average\\delay~ jitter (ms)\end{array}$ & 33.6 & 8.7 & 28.3 & 14 & 9.3 \\
  \hline
\end{tabular}

\end{table}

\begin{table*}
  \centering
  \caption{Some statistics on arrival rates for different AQM}\label{stats3}
\begin{tabular}{|c|c|c|c|c|}
  \hline
   & RED & REM & PI & SF \\
  \hline
  \begin{tabular}{c}
    users \\ Mean (pkt/s) \\Stand. dev. (pkt/s)
    \end{tabular}
    & \!\!\!\! \begin{tabular}{c|c|c}1&2&3\\ 147&83&148 \\61&36&372  \end{tabular}\!\!\!\! & \!\!\!\! \begin{tabular}{c|c|c}1&2&3\\ 147&81&60 \\56&34&23  \end{tabular}\!\!\!\! & \!\!\!\! \begin{tabular}{c|c|c}1&2&3\\ 146&88&109 \\66&35&335 \end{tabular}\!\!\!\! &\!\!\!\! \begin{tabular}{c|c|c}1&2&3\\ 104&91&88 \\41&30&40  \end{tabular} \!\!\!\! \\
  \hline
\!\! Jain's fairness index\!\! & 0.9450 & 0.8703 & 0.9579 & 0.9946 \\
  \hline
\end{tabular}

\end{table*}

\section{CONCLUSION}
In this paper the design of an AQM for congeston control of a single router has been presented. The considered topology consists in several TCP sources sending long-lived flows through a router to their respective receivers.To supply the TCP congestion control mechanism, an AQM must be implemented to the router. Based on a modified mathematical model of the protocol behavior, a such AQM has been developed with control theory tools. Indeed, in a time delay system framework, a control law has been proposed and then the stability analysis of the feedback system has been performed. Consequently, the regulation of flows and the queue size of at the router is ensured. At last, a numerical example and simulations have shown the effectiveness of the proposed methodology.

\bibliographystyle{plain}
\bibliography{mabiblio}

\end{document}